\documentclass[aps,twocolumn,superscriptaddress,showpacs,amssymb,prb,floatfix,preprintnumbers]{revtex4-1}
\usepackage{graphicx,color}
\usepackage{amssymb}

\newcommand{\slrr}      {$T_1^{-1}$}

\newcommand{\as}        {$^{75}$As}

\newcommand{\tn}     {$T_{\rm N}$}

\newcommand{\slrrtext}  {spin-lattice-relaxation rate}

\newcommand{\bafenias}     {Ba(Fe$_{1-x}$Ni$_x$)$_2$As$_2$}
\newcommand{\bafecoas}     {Ba(Fe$_{1-x}$Co$_x$)$_2$As$_2$}

\newcommand{\bafeas}     {BaFe$_2$As$_2$}

\begin{document}

\thispagestyle{myheadings}

\title{Local Magnetic Inhomogeneities in Lightly Doped BaFe$_2$As$_2$}

\author{A. P. Dioguardi}
\author{N. apRoberts-Warren}
\author{A. C. Shockley}
\affiliation{Department of Physics, University of California, Davis, CA 95616}

\author{S. L. Bud'ko}
\author{N. Ni}
\author{P. C. Canfield}
\affiliation{Ames Laboratory U.S. DOE and Department of Physics and Astronomy, Iowa State University, Ames, Iowa 50011, USA}

\author{N. J. Curro}
\email{curro@physics.ucdavis.edu}
\affiliation{Department of Physics, University of California, Davis, CA 95616}


\date{\today}

\begin{abstract}
We report \as\ NMR measurements in \bafeas\ doped with Ni.  Like Co, Ni doping suppresses the antiferromagnetic and structural phase transitions and gives rise to superconductivity for sufficiently large Ni doping. The spin lattice relaxation rate diverges at \tn, with a critical exponent consistent with 3D ordering of local moments. In the ordered state the spectra quickly broaden inhomogeneously with doping. We extract the average size of the ordered moment as a function of doping, and show that a model in which the order remains commensurate but with local amplitude variations in the vicinity of the dopant fully explains our observations.

\end{abstract}


\maketitle

The recent discovery of the iron arsenide superconductors has reignited interest in the interplay of superconductivity and magnetism in condensed matter.\cite{LaOFFeAsNature,BaFe2As2discovery}  There are several iron arsenide families that exhibit superconductivity either under pressure or with chemical doping.\cite{Kimber2009}  Each family, however, contains a common structural element consisting of FeAs planes, in which the Fe 3d orbitals hybridize with the As p orbitals, giving rise to multiple bands that cross the Fermi energy.\cite{Singh2008PRL,mazinPnictidesProblems}  In the parent state (either undoped or at ambient pressure) the nesting of two of these Fermi surfaces gives rise to a spin density wave instability. \cite{YinPickett} It is believed that doping (both in-plane and out-of plane) and/or pressure tunes the chemical potential and modifies the nesting condition, which consequently suppresses the long range antiferromagnetic order. For sufficiently large electron or hole doping superconductivity emerges.\cite{doping122review}  Although it appears that the superconductivity may be intimately related to the antiferromagnetic fluctuations present in these systems, details of how these fluctuations emerge and are related to the superconductivity remain clouded. \cite{MazinSchmalian2009}  Furthermore, it is unclear how the superconducting condensate can remain intact in the presence of a high concentration of in-plane impurities.\cite{balatskyRMP}  Detailed experimental studies of the local effects of dopants therefore are crucial to understand the physics that drive the evolution of long range order in these materials.

In order to shed light on the influence of dopants on the antiferromagnetic order, we have conducted \as\ Nuclear Magnetic Resonance (NMR) studies in a series of Ni doped \bafenias\ crystals. The advantage of this particular system is that large high quality single crystals are available, and Ni has the greatest effect of any transition metal dopants so that superconductivity is reached at only $\sim$2\%. \cite{doping122review,doping122review,NiCanfield122review}  We find that the As NMR spectrum in the antiferromagnetic phase broadens inhomogeneously very quickly with doping, reflecting a large distribution of local hyperfine fields.  This distribution can be understood by realizing that the local spin polarization surrounding the dopant site is not only reduced on-site (at the Ni), but also extends to several of the surrounding Fe sites.  In the paramagnetic state, we find that the nuclear \slrrtext\ reflects a continuous evolution of three dimensional critical spin fluctuations associated with the phase transition at $T_N(x)$.

\begin{figure}
\includegraphics[width=1.0\linewidth]{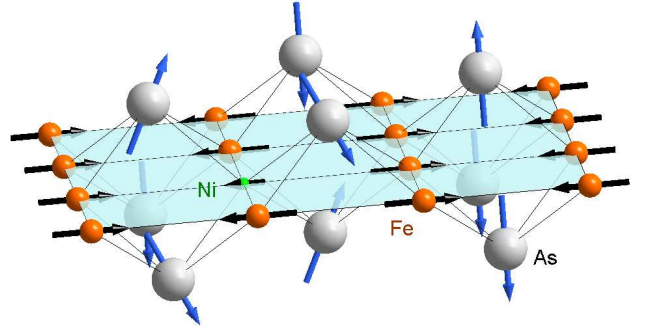}
\caption{\label{fig:HFstructure} (color online) The local structure of the hyperfine fields at the As surrounding a Ni impurity.  Black (blue) arrows indicate the direction of the ordered moment (As hyperfine field).   The hyperfine field at the nearest neighbor As sites acquires components in the $ab$ plane. }
\vspace{-0.15in}
\end{figure}

Single crystals of \bafenias\ were grown by the self-flux method as described in \cite{doping122review,NiCanfield122review}.  The Ni concentrations were determined via microprobe analysis.\cite{doping122review,NiCanfield122review}  NMR spectra were obtained by integrating the spin echo as a function of applied external field oriented along the crystal $c$ direction at constant frequency $f = 48.28$ MHz.  There is a single As site per unit cell located symmetrically between four nearest neighbor Fe atoms, with staggered displacements along the $c$-axis (see Fig. \ref{fig:HFstructure}).  \as\ has spin $I=3/2$, and the resonant condition is given by the nuclear spin Hamiltonian:
\begin{equation}
\label{eq:hamiltonian}
\mathcal{H} = \gamma\hbar\hat{I}_zH_0 + \frac{h\nu_{cc}}{6}\left[3\hat{I}_z^2-\hat{I}^2 - \eta(\hat{I}_x^2 - \hat{I}_y^2)\right] + \mathcal{H}_{\rm hf},
\end{equation}
where $\gamma=0.7292$ kHz/G is the gyromagnetic ratio, $H_0$ is the external applied field, $\hat{I}_{\alpha}$ are the nuclear spin operators, $\nu_{cc}$ is the component of the electric field gradient (EFG) tensor along the $c$-direction, $\eta$ is the asymmetry parameter of the EFG tensor, and $\mathcal{H}_{\rm hf}$ is the hyperfine interaction between the As nuclear spins and the Fe electron spins. The hyperfine coupling is given by:
\begin{equation}
\label{eq:hyperfine}
\mathcal{H}_{\rm hf} = \gamma\hbar\mathbf{\hat{I}}\cdot\sum_{i\in nn}\mathbb{B}_i\cdot\mathbf{S}(\mathbf{r}_i),
\end{equation}
where the sum is over the four nearest neighbor Fe spins $\mathbf{S}(\mathbf{r}_i)$, and the components of the hyperfine tensor $\mathbb{B}$ are given by: $B_{aa} = B_{bb} = 6.6$ kOe/$\mu_B$, $B_{cc} = 4.7$ kOe/$\mu_B$  and $B_{ac} = 4.3$ kOe/$\mu_B$ as determined in the parent compound by Kitagawa \textit{et al}.\cite{takigawa2008}

\begin{figure}
\includegraphics[width=1.0\linewidth]{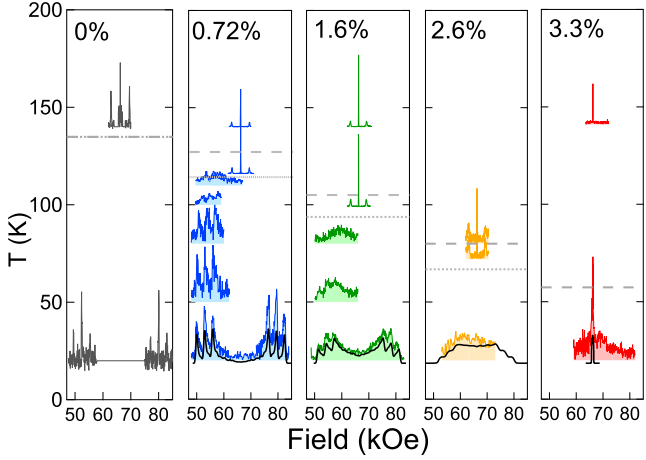}
\caption{\label{fig:spectra} (color online) Field-swept spectra of the \as\ in \bafenias\ at constant frequency of 48.28 MHz for various dopings and temperatures.  The data for $x=2.6$\% at 20K has been shifted down by 10 K for comparison with other dopings. Gray dashed (dotted) lines indicate $T_s$ ($T_N$).  The solid black lines are simulations as described in the text, for $\alpha = 0.6$ and $\lambda = 2.5$a.}
\vspace{-0.15in}
\end{figure}

Fig. \ref{fig:spectra} shows field-swept spectra for the As for several different dopings as a function of temperature.  For sufficiently high temperatures in the paramagnetic phase \bafenias\ has tetragonal symmetry (space group $I4/mmm$), and hence the EFG at the As site has axial symmetry with $\eta=0$.  The spectra consist of three resonances at fields $H_0=(f-n\nu_{cc})/\gamma$, where $n=-1$,0, or 1, and $\nu_{cc} \approx 2.5$ MHz.   $\nu_{cc}$ is strongly temperature dependent,\cite{takigawa2008} but surprisingly we find little or no change in the average value of $\nu_{cc}$ with doping. Rather, the spectra reveal an increasing quadrupolar satellite ($n=\pm1$) linewidth, reflecting an distribution of local EFGs as expected in the presence of dopants.  By $x=3.3$\% the disorder-induced broadening is so large that we are no longer able to discern the quadrupolar satellites.  The increased quadrupolar linewidth, however, is minor compared with the large distribution of hyperfine fields in the antiferromagnetic state.

Below a temperature  $T_s(x)$ the system undergoes a structural distortion to an orthorhombic phase (space group $Fmmm$).   In the parent compound $T_s(0) = 134$ K  coincides with $T_N$ and the sudden appearance of antiferromagnetic order. Upon doping, however, this first order phase transition quickly becomes second order, and the two transition temperatures $T_s(x)>T_N(x)$ separate.\cite{doping122review,NiCanfield122review}  These transitions are reflected in the NMR spectra by a change in the EFG at $T_s(x)$ as well as the onset of static internal hyperfine fields below $T_N(x)$ that shift the spectra both to higher and lower resonance fields (Fig. \ref{fig:spectra}).   In the presence of the internal field $\mathbf{H}_{\rm int}$ the resonance condition becomes $f = \gamma|\mathbf{H}_0 + \mathbf{H}_{\rm int}| + n\nu_{cc}$. Since  $\mathbf{H}_{\rm int}$ is either parallel or antiparallel to $\mathbf{H}_0$ ($||~\hat{c}$) there will be six resonances at: $H_0 = (f-n\nu_{cc})/\gamma \pm H_{\rm int}$, as seen in Fig.\ref{fig:spectra} for undoped BaFe$_2$As$_2$  below \tn.   We find $H_{\rm int} = 14.95$ kOe, in agreement with previous work. \cite{takigawa2008}   This observation is consistent with the hyperfine coupling (Eq. \ref{eq:hyperfine}) for magnetic ordering given by $\mathbf{Q} = (\frac{\pi}{a},0,0)$ with moments $S_0 = 0.87\mu_B$ along the $(100)$ direction, as observed by neutron scattering (see Fig. \ref{fig:HFstructure}).\cite{Goldman2008PRB}  It is crucial to note that even though the moments lie along the $(100)$ direction, the hyperfine field lies along $(001)$.  The reason is that the As is located symmetrically between four nearest neighbor Fe atoms so that the components of $\mathbf{H}_{\rm int}$ in the $ab$ plane cancel out by symmetry.  A similar effect is at play for the planar oxygen site in the cuprates,\cite{MMPT1inYBCO} except that in the iron arsenides the fact that the As is not in the plane of the Fe atoms gives rise to the $c$-axis component of the hyperfine field.

\begin{figure}
\includegraphics[width=1.0\linewidth]{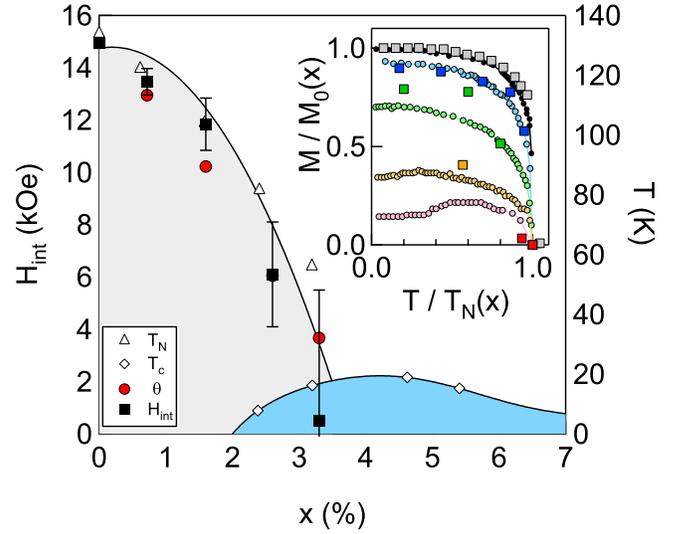}
\caption{\label{fig:phasediagram} (color online) $H_{\rm int}$ (left axis), $T_N$ and $T_c$ (right axis) as functions of doping in  \bafenias.  The open points are determined by bulk measurements \cite{canfield3ddoping} and the solid lines are guides to the eye.  INSET:  The reduced moment $M/M_0$ versus reduced temperature $T/T_N$ as measured by NMR in \bafenias\ ($\square$) and by elastic neutron scattering  in \bafecoas\ ($\circ$).\cite{FernandesNS122} The colors correspond to the same dopings as in Fig. \ref{fig:spectra}.}
\vspace{-0.15in}
\end{figure}

Upon doping, the internal field is reduced and the spectra broaden dramatically in the ordered state.  The spectra are best characterized by a distribution of $\mathbf{H}_{\rm int}$. The average internal field $\langle{H}_{\rm int}\rangle$ (determined by the peak of the spectra)  tracks $T_N(x)$  as a function of doping and is shown in Fig. \ref{fig:phasediagram}. Indeed, $H_{\rm int}~\sim~S_0$, is proportional to the sublattice magnetization, $S_0$, and is a direct probe of the antiferromagnetic order parameter.  The reduced internal field, $H_{\rm int}(x,T)/H_{\rm int}(0,T)$, is shown in the inset.  For comparison, recent  elastic neutron scattering measurements on Co doped \bafecoas\ are shown as well.\cite{FernandesNS122} The similar trends in the doping and reduced temperature matches those of the phase diagrams of these two systems.\cite{doping122review,NiCanfield122review}  For sufficiently large dopings ($x\gtrsim2.6$\%) the width of the internal field distribution becomes so large that the spectral features are washed out.   Similar effects were observed in \bafecoas, and attributed to the presence of incommensurate magnetic order.\cite{Laplace2009,ImaiLightlyDoped}  Although the spectra reveal a broad distribution of $\mathbf{H}_{\rm int}$, it is unclear whether there is an incoherent distribution of local Fe spin polarizations, $\mathbf{S}(\mathbf{r})$, or an incommensurate modulation.  Both effects would give rise to a distribution of internal fields consistent with the spectra.   However, in the strongly correlated antiferromagnetic state a local impurity (Ni) can give rise to dramatic changes to the local spin structure even in the presence of commensurate order. Kemper and collaborators have calculated the local spin density in the vicinity of a dopant atom in the antiferromagnetic state of the BaFe$_2$As$_2$ system using density function theory (DFT), and find that the polarization of the neighboring Fe are strongly modified as result of the impurity.\cite{KemperHirschfeld}  Since the hyperfine interaction (Eq. \ref{eq:hyperfine}) couples each As nucleus to four Fe sites, small changes to the spin of any Fe site can alter the hyperfine field of several As nuclei.   If one of the Fe spins were slightly reduced due to its proximity to a dopant Ni (or Co) atom, then the hyperfine field at several neighboring As sites would acquire a component in the $ab$ plane (see Fig. \ref{fig:HFstructure}) and the resonant fields would be given by:
\begin{equation}
\label{eq:Hres}
H_0 = \sqrt{\frac{1}{\gamma^2}(f-n\nu_{cc})^2-(H_{\rm int}^{ab})^2}-H_{\rm int}^{c}.
\end{equation}
Clearly, a distribution of either $H_{\rm int}^{ab}$ or $H_{\rm int}^{c}$ could broaden the NMR spectra significantly.

In order to quantify the effect of perturbations of the local hyperfine field in the vicinity of a dopant atom, we have simulated the spectra using a simple model designed to mimic the spin density observed in DFT calculations.  We start with a 2D square lattice of $100\times 100$ Fe spins with impurities located randomly at positions $\mathbf{r}_{\rm imp}$.  The spin moments are given by $\mathbf{S}(\mathbf{r}) = \mathbf{S}_0\Phi_x(\mathbf{r}) \cos(\mathbf{Q}\cdot\mathbf{r})$, where $\Phi_x(\mathbf{r}) = 1-\alpha \sum_i\exp{\left[ -(\mathbf{r}-\mathbf{r}_{\rm i})^2/2\lambda^2 \right]}$ represents the reduction of the local spin polarization in the vicinity of a dopant.   $\alpha$ is a constant that represents the suppression of spin density magnitude at an impurity site, and $\lambda$ is a constant that represents the spatial extent of the suppression.  This model captures the relevant features of the DFT calculations, which indicate that not only is the impurity site spin moment renormalized, but the magnetization over several neighboring sites in the Fe plane is also reduced.\cite{KemperHirschfeld}   Given this spin density profile, we calculate the hyperfine fields at each As site in the lattice using Eq. \ref{eq:hyperfine}, and determine the histogram of resonant fields using Eq. \ref{eq:Hres}.  These histograms are shown in Fig. \ref{fig:spectra} for $\alpha = 0.6$ and $\lambda = 2.5a$.  The simulated spectra clearly exhibit the same general trend versus doping as our experimental spectra, and the values of $\alpha$ and $\xi$ are reasonably close to the DFT calculations, suggesting that the origin of the broad NMR spectra is indeed natural consequence of doping-induced disorder in the antiferromagnetic state rather than a change from commensurate to incommensurate ordering.

It is important to note that this non-local effect of the Ni dopants is a consequence of the strong correlations present in the long-range ordered antiferromagnetic state.  In the paramagnetic state, NMR experiments show little or no change of the spin fluctuations at As sites that are nearest neighbors to the dopant atoms.\cite{imaiBa122overdoped}  In this case,  the magnetic correlations are not sufficiently large to induce the non-local effects seen in other strongly correlated systems.\cite{AlloulHirschfeld,CeCoIn5CdDroplets}

\begin{figure}
\includegraphics[width=1.0\linewidth]{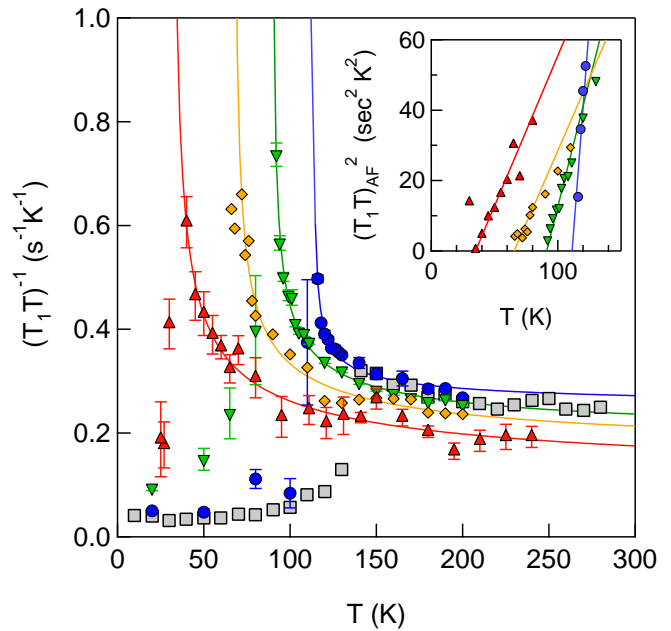}
\caption{\label{fig:T1Tinv} (color online) $(T_1T)^{-1}$ versus temperature in \bafenias\ for $x=0$ ($\blacksquare$, gray, reproduced from \cite{takigawa2008}), $x=0.72$\% ($\bullet$, blue), $x=1.6$\% ($\blacktriangledown$, green), $x=2.6$\% ($\blacklozenge$, yellow), and $x=3.3$\% ($\blacktriangle$, red). Solid lines are fits as described in the text. INSET: $(T_1T)^2_{\rm AF}$ versus $T$ using the same symbols as in the main figure. Solid lines are linear fits indicating behavior characteristic of 3D fluctuations.}
\vspace{-0.15in}
\end{figure}

In order to investigate the evolution of the spin dynamics with doping, we measured the spin lattice relaxation rate, \slrr, at the central transition ($I_z=-\frac{1}{2}\leftrightarrow \frac{1}{2}$).   The data are well fit to a single $T_1$ component, and the temperature and doping dependence are shown in Fig. \ref{fig:T1Tinv}.  $(T_1T)^{-1}$ probes the dynamical spin susceptibility of the Fe spins, and is given by:
\begin{equation}
\frac{1}{T_1T} = \gamma^2k_B\lim_{\omega\rightarrow 0}\sum_{\mathbf{q},\beta}F^2_{\beta}(\mathbf{q})\frac{\chi_{\beta}''(\mathbf{q},\omega)}{\hbar\omega},
\end{equation}
where the form factor $F(\mathbf{q}$) is the Fourier transform of the hyperfine coupling and $\chi_{\beta}''(\mathbf{q},\omega)$ is the dynamical susceptibility \cite{Moriya1974}.
We find that $(T_1T)^{-1}$ exhibits a dramatic enhancement in the paramagnetic state just above $T_N$ due to critical slowing down of the spin fluctuations.\cite{MoriyaT1formula}  For $x=0$, $(T_1T)^{-1}$ exhibits a discontinuity at $T_N$, consistent with the first order nature of the phase transition.  Under doping, however, the transition becomes second order.\cite{BaekBaFe2As2PRB2008} The data are well fit to $(T_1T)^{-1} = a + b/(T-T_N)^{1/2}$, where the first term represents a Korringa relaxation by itinerant quasiparticles, and the second term arises from 3D fluctuations of nearly antiferromagnetic local moments.\cite{Moriya1974}   $T_N$ is plotted in Fig. \ref{fig:phasediagram}, and agrees well with bulk measurements.\cite{doping122review,NiCanfield122review}  The inset of Fig. \ref{fig:T1Tinv} shows $(T_1T)^2_{\rm AF}\equiv (1/(T_1T)^1 - a)^{-2}~\sim~\xi_{\rm AF}^{-1}$, where $\xi_{\rm AF}$ is the antiferromagnetic correlation length.  Clearly $\xi_{\rm AF}~\sim~(T-T_N)^{-\nu}$ with $\nu=1/2$, the mean field result for 3D fluctuations.
Upon doping, these critical fluctuations are suppressed as $T_N(x)$ decreases.  Similar trends were observed in \bafecoas, and attributed to the evolution of both inter- and intra-band scattering.\cite{imaiBa122overdoped} For sufficiently large electron doping, the hole Fermi surface centered at $\Gamma$ disappears as well as the strong intra-band scattering responsible for the SDW ground state.\cite{SDWtheory}

To summarize, we have measured the spectra of \bafenias\ as a function of doping and temperature in the SDW phase and found a broad distribution of internal hyperfine fields consistent with long-range modulations of the local spin density around the dopant atoms in the ordered state.  These observations are consistent with recent DFT calculations and neutron scattering measurements.\cite{KemperHirschfeld,FernandesNS122} The spin-lattice relaxation rate reveals strong 3D critical spin fluctuations that accompany the suppression of $T_N$ with doping.  Ni doping is qualitatively similar to Co, except that it contributes more electrons per dopant atom.\cite{doping122review,NiCanfield122review}  We find no evidence for low frequency spin fluctuations in the ordered state associated with glassy behavior as observed in the cuprates.\cite{ImaiLightlyDoped,BorsaPRL1993}  Future studies of the region of coexisting antiferromagnetism and superconductivity may shed new light on the question of why the superconducting condensate can survive in the presence of pair breaking scattering centers.

We thank P. Hirschfeld, L. Kemper, S. Balatsky and M. Graf for stimulating discussions. This project was supported by the University of California under grant UCOP-01-09. Work at the Ames Laboratory was supported by the US Department of Energy - Basic Energy Sciences under Contract No. DE-AC02-07CH11358.


\end{document}